# The Orchestral Analog of Molecular Biology

Stan Bumble, Physics Department, Community College of Philadelphia
1700 Spring Garden Street, Philadelphia, PA
sbumble@ccp.edu

Introduction
Signal Processing and EIIP (Electron-Ionization Interaction Potential)
SYNPROPS and Knapsack
Codon Graphlets
QSARs
Multiobjective Search
Regression and Matrix Formulation
Homology Modeling Interactome: Scale-Free or Geometric?
The Orchestral Analog of Molecular Biology
Conclusion
Appendix 1
Appendix 2

**Abstract**

Signal processing (SP) techniques convert DNA and protein sequences into information that lead to successful drug discovery. One must, however, be aware about the difference between information and entropy[1]. Eight other physical properties of DNA and protein segments are suggested for SP analysis other than ones already used in the literature. QSAR formulations of these properties are suggested for ranking the amino acids that maximize efficiency of the amino acids in proteins. Multiobjective programs are suggested for constraining or searching the components of such sequences. Geometric maps of the networks of proteins are preferable to scale-free descriptions in most cases. The genetic code is presented as graphlets which show interesting correspondence to each other, leading to possible new revelations.

**Introduction**

Biochemical reactions are controlled by enzymes or regulators, and these are proteins. Proteins depend on their three-dimensional shape, which depends on the protein's sequence of amino acids. Twenty types of amino acids can occur in proteins. To form a protein, these twenty amino acids are strung together in a precise sequence. It is this sequence which determines the three dimensional shape into which the proteins will fold, as well as its activity. In order to proceed from the genome to function one may need to know the interpretation of the protein structure as to its function. However, since the function may depend on the permutation of these amino acids (with repeats) along the lengthy protein scaffold, this may be practically an NP problem. Fortunately, recently the advent of signal processing methods and information technology are leading to solutions that bode well for the medical profession, the drug discovery process, the quality of life and the longevity of mankind. The quantification of biological networks into expression is also a well nigh NP problem as in physics the network problems that have led to approximate solutions were all regular lattices and the e.g., PPI (protein-protein interactome networks) are irregular webs.

**Signal Processing**[2,3,4,5,6]

Signal processing is not a new technology. For example the IR, microwave and Raman structure of molecules were used to determine their thermodynamic function using statistical mechanics which led to solutions of their complex equilibrium and reaction rate results using computers for many years. Thus spectroscopy, akin to signal processing, led to significant results. The problems that arose with biological molecules is that they can be quite large and the computation of their vibrations, rotations, etc. becomes a very difficult problem indeed. Results using a computer program called Chemkin have shown, however, that the kinetics of small biological molecules are solvable. A program called Therm does find thermodynamic functions for small molecules and can theoretically find such data for large molecules with proper data input. There are encouraging reports that scientists in Japan are to publish thermodynamic functions for biological molecules.

In a book by Cosic, The Resonant Recognition Model of Macromolecular Bioactivity, the electrons along a protein molecule are considered as delocalized and the charges moving along the protein backbone as productive of electromagnetic effects, whose spectral characteristics, that correspond to their energy distribution, correspond to their energy distribution along the protein. A numerical sequence representing each amino acid in the sequence is then associated with a corresponding electron-ion interaction value. Signal processing techniques were then applied to protein (and gene) sequences to search for information embedded in these biological systems. Now signal processing techniques have been applied most everywhere in bioinformatics and will continue to play an important role in the study of biomedical problems. Thus the frequency spectra was described with the discrete Fourier transform (DFT) technique. Comparison of results for groups of proteins, which performed a similar function, showed that all possessed a wave form in common, a single characteristic frequency. Different groups of proteins performing different functions had a different characteristic frequency. Also, receptor target molecules. to which particular proteins bind, had a matching characteristic to the ligand molecule.

The Fourier method discovered "global" periodicities and could not extract hidden localized periodicity very well. These could be useful for underlying construction rules. Such a correlation function could compare each DNA base with its various neighbors. Regular features in DNA sequences were observed after further Fourier and wavelet processing. Significant features of the DNA sequence showed that the noncoding sequences had spectra similar to random sequences, while coding sequences revealed specific periodicities of both variable length and also a common periodicity of three.

Wavelet methods were also used later to represent the spectra of proteins. Furthermore, the spectra were transformed from many peaks to singular characteristic peaks that represented the protein for docking purposes (which is usually used in the process of drug discovery). Further work by others stated that this representation of a protein by one characteristic frequency is not valid and instead should be placed by several characteristic frequencies.

Considering that the Hartree-Fock approximation of quantum mechanics was applied to such a complex problem, and that some literature sources apply the band structure of solid state physics to this problem and that so much emphasis was placed on one property (the EIIP) to be the sole important characterizing property of these proteins; how valid can these working hypotheses be? According to Veljkovic, molecular electrostatic potential provides a basis for the initial assessment of specific biological activity of a molecule. The potential that is created around a molecule is what is "sensed" by another molecule in its vicinity. Such interactions are effective in less than 2 angstroms and is a "short-distance interaction" (SDI). Besides the electrostatic phenomena, the 3-D structures of the molecules are important. Veljkovic believes that the electrostatic potential is also important in the "recognition" processes that enzyme substrates or drug-receptor interactions display. This second interaction, which is termed the "long-distance interaction" (LDI), occurs when the molecule and its receptor are at a relatively large separation (100-1000 angstroms). It directly influences a number of productive collisions between interacting biochemical molecules and their kinetics. This potential involves molecular processes such as receptor-ligand, antibody-antigen and enzyme-substrate kinds. A psuedopotential parameter was used, the electron-ion interaction potential (EIIP), for development of the ISM (Informational Spectrum Model), a virtual spectroscopy method for informational analysis of protein and nucleotide sequences. The signal obtained has been decomposed in periodical functions by Fourier transformation. The result is a series of frequencies and amplitudes. This technique allowed detection of code/frequency pairs, characteristic for interacting molecules, and for determining their specific long-distance recognition and targeting. The ISM has helped elucidate structural motifs with defined physico-chemical interacting molecules, determining their specific long-distance recognition and targeting. ISM has been used for structure/function analysis of protein and DNA sequences, as well as for denovo design of biologically active peptides.

Stambuk et alia state the Cosic model was confirmed by successful prediction of macromolecular receptor binding, enzyme and oncogene activity, protein-DNA interactions and bioactive parts of cytokines, hormones, viral proteins and antibodies. However, the problem of defining its structural classes was not solved by the single characteristic approach. A single common characteristic frequency is often not found due to sequence diversity irrespective of the secondary structure. They then analyzed a limited number of the most prominent frequency peaks in the protein spectrum of different folds. The pattern recognition of protein EIIP frequency periodogram obtained by spectral Fourier analysis enabled classification of the secondary protein folds. A single periodogram peak has a low information content, unable to provide a discriminating parameter for the complex structure, which is highly diverse, and consequently often spectrally deviant, with respect to the species and function. The decision tree based pattern analysis of Fourier periodogram frequency peaks seemed to provide a useful alternative to the common frequency determination, since it enabled accurate recognition of α- and β-protein folding types.

A method[7,8] for condensing the information in multiple alignments of proteins into a mixture of Dirichlet densities over amino distributions exists. Dirichlet mixture densities are designed to be combined with observed amino acid frequencies to form estimates of expected amino acid probabilities at each position in a profile hidden Markov model, or other statistical model. These estimates give a statistical model greater generalization capacity, so that remotely related family members can be more reliably recognized by the method. The method employed to estimate and use Dirichlet mixture priors are based firmly on Bayesian statistics. While biological knowledge has been introduced only indirectly from the multiple alignments used to estimate the mixture parameters, the mixture priors produced agree with accepted biological understanding. The effectiveness of Dirichlet mixtures for increasing the ability of statistical models to recognize homologous sequences has been demonstrated experimentally.

## SYNPROPS and KNAPSACK[9,10, 11, 12]

In books written by the author, two programs are introduced. One is called SYNPROPS, the other is called KNAPSACK. The former obtains physical properties for molecules using chemical groups that have been obtained from experimental data. This program has been used to approximate physical properties for amino acids and also for segments of proteins containing amino acids. The later program determines the frequency of fifteen of the amino acids contained in a protein of given length with constraints on the physical properties of the protein. This was done for the physical properties studied[9,10]. One objective was to see how the properties studied before were related to the EIIP.

SYNPROPS uses the data and formulas of Cramer that are inserted into a spread sheet program such that once the composition of molecules is given in terms of molecular groups, the properties of designated molecules can be ascertained to a given level of accuracy. Cramer derived the data originally for the molecules with data from the handbooks of chemistry and physics and the predictions of properties of other molecules were deemed to be accurate and the statistics used in the whole process were above suspicion. The SYNPROPS matrix of data and formulas were also used in the usual optimizing programs contained in the PC spreadsheet programs of Quattro Pro, Excel, etc., so that one could find the best structures of molecules (and mixtures of molecules) for desired properties, and furthermore, the whole procedure was reduced to one where methods of matrix algebra could be utilized to proceed backwards and forwards in the manipulation of properties and structures of molecules. These methods were used for determining properties for biological molecules, such as amino acids and proteins and in applications such as environmental substitutions, toxicity and drug discovery.

Table 1

| residue | dipole | -logparc | V/20 | hydrophob | logactco | sol par | CHI | mol refract | EIIP |
|---|---|---|---|---|---|---|---|---|---|
| ala | 1.63945 | 2.8998 | 3.74845 | -2.60426 | 6.577995 | 8.27661 | 35.3748 | 17.11992 | 1.865 |
| asn | 4.67705 | 5.21377 | 4.43545 | -1.44596 | 14.93858 | 16.16167 | 13.8594 | 25.02192 | 0.18 |
| asp | 3.07055 | 4.40374 | 4.27155 | -7.37356 | 12.51525 | 13.07255 | 21.6879 | 23.64277 | 6.315 |
| cys | 2.38845 | 3.32977 | 4.2553 | -4.04259 | 9.367475 | 10.638 | 30.5513 | 23.50322 | 8.29 |
| gln | 4.56335 | 4.69192 | 5.2949 | -0.94134 | 14.83021 | 15.962 | 16.3845 | 29.53745 | 3.805 |
| glu | 2.95685 | 3.88189 | 5.131 | -6.86894 | 12.40688 | 12.872 | 24.2147 | 28.1583 | 0.29 |
| gly | 1.77355 | 3.42677 | 2.88915 | -2.93481 | 6.678975 | 8.48389 | 32.8367 | 12.61641 | 0.5 |
| iso | 1.34775 | 1.46208 | 6.456 | -0.4382 | 6.200525 | 7.39963 | 42.7831 | 30.72551 | 0 |
| leu | 1.34775 | 1.46208 | 6.456 | -0.4382 | 6.200525 | 7.39963 | 42.7831 | 30.72551 | 0 |
| lys | 2.06725 | 3.24955 | 6.4425 | -7.57275 | 10.6585 | 9.65799 | 32.0947 | 34.65878 | 1.855 |
| met | 2.31305 | 2.62797 | 6.00895 | 1.53475 | 8.950935 | 9.48157 | 35.9034 | 33.5721 | 8.23 |
| phe | 1.77705 | 1.47151 | 7.0071 | -3.35187 | 8.556295 | 9.13923 | 41.7417 | 41.19408 | 4.73 |
| ser | 2.88225 | 4.72784 | 3.4832 | -7.33078 | 11.63029 | 12.25083 | 21.2775 | 17.85461 | 8.29 |
| thr | 2.81825 | 4.333789 | 4.47185 | 6.17395 | 11.46956 | 11.77301 | 23.6356 | 2.42914 | 4.705 |
| val | 1.46145 | 1.98392 | 5.59655 | 0.94282 | 6.308895 | 7.59921 | 40.258 | 26.20998 | 2.85 |

KNAPSACK is a program adopted from discrete mathematics and also is processed on a spreadsheet such as Excel or Quattro Pro. It calculates the optimum distribution of objects that can be contained in a chromosome, if the property values for each object can vary to a different degree. Originally, it calculated the frequency of each amino acid in a protein if the protein was limited to a maximum number of amino acids. Then it was also extended to also determine the practical appropriate sequence of these amino acids in the protein. The number of amino acids considered was fifteen instead of twenty because the data to regress the properties of the five remaining amino acids with their structure was not available at the time but this can be addressed in the future. The KNAPSACK program utilizes the variables $n_i$ for the number of each of the amino acid residue i and Pi for the particular property for the amino acid residue i. Then the summation Sum (i = 1 to i = 15) for the function $S = n_i P_i$ from i = 1 to i = 15 is maximized or minimized according to what particular property Pi represents. Now the summation of N =$n_i$ (from i = 1 to i = 15) is fixed according to the length of the peptide or the protein chain. This means the extreme of the of sum of $S = n_iPi$ is found under the constraint that the sum of ni (N) is a fixed number. When this is done and the values of $n_i$ are allowed to vary while the extremum of $S = n_iP_i$ is found, under the constraint that the sum of $n_i$ is a fixed value, then the spectrum of the values of each of the variables $n_i$ is found (where each of these values is the number of each particular amino acid residue found). This is then the frequency spectrum of amino acids under the conditions that were set.

Knapsack uses the physical properties in the above table, eight of which come from SYNPROPS and that for EIIP from the literature for the fifteen amino acids. The total number of amino acids is fixed to any number desired. A property or function of any or all of the properties can be maximized or minimized, Each number of the amino acids in the result is constrained to be an integer and any number of them can be constrained to be above, below, equal to or between any set of numbers desired.

**Codon Graphlets**

The genetic code is shown in Table 2.

**Table 2**

|   | T | C | A | G |
|---|---|---|---|---|
| **T** | TTT Phe (F)<br>TTC "<br>TTA Leu (L)<br>TTG " | TCT Ser (S)<br>TCC "<br>TCA "<br>TCG " | TAT Tyr (Y)<br>TAC<br>TAA Ter<br>TAG Ter | TGT Cys (C)<br>TGC<br>TGA Ter<br>TGG Trp (W) |
| **C** | CTT Leu (L)<br>CTC "<br>CTA "<br>CTG " | CCT Pro (P)<br>CCC "<br>CCA "<br>CCG " | CAT His (H)<br>CAC "<br>CAA Gln (Q)<br>CAG " | CGT Arg (R)<br>CGC "<br>CGA "<br>CGG " |
| **A** | ATT Ile (I)<br>ATC "<br>ATA "<br>ATG Met (M) | ACT Thr (T)<br>ACC "<br>ACA "<br>ACG " | AAT Asn (N)<br>AAC "<br>AAA Lys (K)<br>AAG " | AGT Ser (S)<br>AGC "<br>AGA Arg (R)<br>AGG " |
| **G** | GTT Val (V)<br>GTC "<br>GTA "<br>GTG " | GCT Ala (A)<br>GCC "<br>GCA "<br>GCG " | GAT Asp (D)<br>GAC "<br>GAA Glu (E)<br>GAG " | GGT Gly (G)<br>GGC "<br>GGA "<br>GGG " |

You may notice this is like a matrix of symbols. There are three sets of boxes: 1) on the diagonal are four boxes of codons for particular amino acids, phe, pro, lys, gly,.that have three identical letters (triplets), 2) the second set consists of six boxes containing codons with the same amino acid, leu, val, ser, thr, ala, and arg, The last set contains codons including the amino acids cys and ser, the amino acids with the highest EIIP, in addition to phe, leu, iso, met, tyr, his, gln, asp, glu, trp, and arg. A codon can be represented as a triangle with the symbols of the nucleotide bases at the vertices. The ones that are triplets can form a graphlet with four triangles where three of them have common bases (or bonds) with the central triangle (a triplet). Other codons can form a graphlet that is a rhombus composed of two triangles with a common base. The internal regularity displayed by the genetic code has been recognized to include physico chemical property correlation, biosynthetic relation of amino acids and the non-random pattern of the genetic code. It is interesting to note that KNAPSACK selected the following amino acids as most favorable for the properties listed as: *asn;* partition coefficient, activity coefficient, solubility parameter, dipole moment, *leu;* CHI, *asp*; EIIP, lys; hydrophobicity, *gly*; volume, and *phe*; molar refractivity.

In an early paper[13, 14, 15, 16] a regular triangular lattice was broken up into subfigures. The subfigures were related by normalization, consistency and equilibrium equations. The latter relations were derived from statistical mechanics and minimization of free energy. In the more complicated network above, we may see a way to follow a similar procedure to treat not only biomolecular networks, but also ligands of proteins docked onto DNA and onto other proteins as well.

**QSARs**

Regression in grossly undetermined systems has emerged as an important means for understanding molecular activity via Quantitative Structure-Activity Relationships (QSAR). These model the relationship activities and physiochemical properties of a set of compounds and they are fundamental to many aspects of the drug design process. The information from a QSAR approach can provide an explanation for the observed differences in binding affinity. It can also provide a quantitative method for the design and prediction of novel potential drug molecules. The methods described here predict physical characteristics of molecules for which no experimental data are available. These can then be used to improve the model of how a drug interacts with its target. QSAR studies can also be applied to near-infrared spectroscopy.

**Multiobjective Search**

Multiobjective optimization is also very useful for such a purpose. Here an objective function is used that encodes all of the desired selection criteria and then a simulated annealing or evolutionary approach is added to identify the optimal (or nearly optimal) subset from among a vast number of possibilities. Many design criteria can be accommodated; those that may be diverse or similar to known actives, etc., and the predicted and/or selectively determined molecules of merit found.

**Regression and Matrix Formulation**

An equation was used for each amino acid:

$\Sigma_i a_i P_i = P_{[EIIP]}$

where i varied from 1 to 8, signifying the eight properties studied before. Then the matrix equation was used:

$A P = P_{[EIIP]}$

where the matrix A represents the matrix of coefficients in the above equations and P the respective numerical properties. Then

$A^{-1} A P = A^{-1} P_{[EIIP]}$ or $P = A^{-1} P_{[EIIP]}$

EIIP=$a_1$(Dipole Moment)+$a_2$(Ln Activity Coefficient)+$a_3$(Volume/20)+$a_4$(Molar Refractivity)-$a_5$(Ln Partition Coefficient)+$a_6$(Hydrophobicity)+$a_7$(Solubility Parameter)+$a_8$(CHI)

The ratios of the coefficients were $a_1/a_2/a_3/a_4/a_5/a_6/a_7/a_8 = 1/.33/.13/.06/.01/.01/.00$ so the last two terms for the properties CHI and solubility parameter could be neglected. This was because their property was assimilated in the other properties and their contributions were minor. In deriving the inverse of a matrix the original matrix must be square. The resulting equation was

EIIP=$a_1$(Dipole Moment)+$a_2$(Ln Activity Coefficient)+$a_3$(Volume/20)+$a_4$(Molar Refractivity)-$a_5$(Ln Partition Coefficient)+$a_6$(Hydrophobicity)

In order to solve the matrix equations above the functions of finding the inverse of the matrix A and the function of matrix multiply functions of Excel and Quattro Pro were used.

It was found that the above equation could be solved, that is, that the property EIIP was a linear function of the six properties designated in the above equation. Thus the single frequency found above is a compendium of such other properties but the use of more than one frequency can be validated in more detail. In using KNAPSACK, the maximum functions of each of the properties were pointing to asn, asn, gly, phe, asn, and lys, resp., for the first through the sixth optimum property as can be seen from the Table 1, but the property of EIIP was a mixture of cys and ser, respectively. When the equation was optimized for the best amino acid overall, the amino acid serine was found with the coefficients used.

**Homology Modeling Interactome: Scale-Free or Geometric?[17]**

A random geometric model provides a more accurate model of the PPI data than the currently popular scale-free model. In the random geometric model the term graphlet is used to denote a connected network with a small number of nodes. The graphlet frequency is the number of occurrences of a graphlet in a network and it is a parameter that shows that the PPI networks are closest to geometric random graphs. The relative frequency of graphlets $N_i(G)/T(G)$ characterizes PPI networks and the networks chosen to model them. Here $N_i(G)$ is the total number of graphlets of type i in a network G and $T(G) = \Sigma N_i(G)$ is the total number of graphlets of G. The similarity between two graphs should be independent of the total number of edges, and depend only upon the differences between the relative frequencies of graphlets. The relative graphlet frequency distance D(G,H) between two graphs G and H is defined as $D(G,H)=\Sigma|F_i(G)-F_i(H)|$, where $F_i(G)=-\log(N_i(G)/T(G))$. Then the distances between several real-world PPI networks and the corresponding SF and GEO random networks were computed and the GEO random networks fit the data an order of magnitude better in the higher-confidence PPI networks and less so in the more noisier PPI networks, except for a fruitfly PPI network with lower confidence interactions. The analysis of the diameters and clustering coefficients further solidifies these results. The fact that the PPI itself may have been made with questionable data[18] and that a protein may have more than one function[19] is pointed out in recent literature. The geometric protein-protein interactome bears a striking resemblance to the order disorder treatment of lattices that occurred in the literature many years ago[13].

**The Orchestral Analog of Molecular Biology[21]**

We have sheet music for solo instruments and orchestra. Each instrument or section of an orchestra has staffs in the score. It now is apparent that the properties are like musical notes and the amino acids are like the instruments or sections of an orchestra in polyphonic music and the lines in the staffs are the sorted properties expressed as if they are the pitch levels in a musical composition. Now the sections in an orchestra may be composed of many instruments and so is DNA. There are nucleotides between the genes whose true meaning or expression are yet to be understood. A musical composition emerges from musical notes in time and the graphlets or motifs are connected in pathways and networks that yields melodies and all the other creations in music as does the DNA and proteins that engage in the process of life. We cannot do signal processing on one property alone because life is more complex. It must be considered as a "polyphonic" analog of music. This can be done by using tables such as that shown as Table 1. Each line of the staff represents the level of a property that, taken together for all the lines and staffs, expresses the composite of the property characteristics. Such information can be derived from a plot of sequences that are observed in laboratories from organisms. It is hoped that the numbers derived for the properties are accurate in the future. Care must be undertaken by scientists to obtain such numbers. For the moment I hope that "it is more important to be consistent than accurate in thermodynamics"[22].

**Conclusions**

Advantage of the signal processing techniques should be extended to other properties than have been used before to provide information useful for disease examination and drug discovery. The use of QSAR (quantitative structure activity relations) will also be useful in such studies. Multiobjective goals are to be more frequent and the programs of SYNPROPS and KNAPSACK can be useful in uncovering the relation of properties, functions and sequences of DNA and proteins. Geometrical networks must be used more often and protein-protein interactomes are to be carefully constructed from valid data. More information can be found from careful analysis of the genetic code. The internal regularity displayed by the genetic code has been recognized to include physicochemical property correlation, biosynthetic relation of amino acids and the non-random pattern of the genetic code[20]. Although the work above has only been carried out with fifteen amino acids, it will be worth the effort to extend this work to all twenty amino acids. This can be accomplished extending the programs, THERM and SYNPROPS and also it is hoped from work to be published from RIKEN in Japan.

**Appendix 1**

As an example of the knapsack results, when the following properties are maximized: dipole moment, hydrophobicity, partition coefficient, EIIP, activity coefficient and molar refractivity and the volume is minimized we find that the program KNAPSACK finds the following results when, at the same time, the number of amino acids is set at 1000, the number of each amino acid is an integer and greater than zero:

|  | dipole moment | hydrophobicity | partition coefficient | volume | EIIP | activity coefficient | molar refraction/5 |
|---|---|---|---|---|---|---|---|
| maximum value | 4677 | -7573 | -5213 | 57783 | 8290 | 14939 | 8239 |
| constraint | =4500 | =6000 | =-5000 | =60000 | =8000 | =14000 | =8000 |
| asn | 790 |  | 692 |  |  | 611 |  |
| asp | 32 | 417 | 61 |  | 1000 | 66 |  |
| cys |  | 38 |  |  |  | 14 |  |
| gly |  | 3 |  | 813 |  |  |  |
| phe |  | 16 |  |  |  |  | 843 |
| ser | 21 | 142 | 85 | 187 |  | 51 |  |
| met |  |  |  |  |  | 7 | 157 |
| gln | 114 |  | 83 |  |  | 104 |  |
| glu | 25 | 128 | 23 |  |  | 64 |  |
| lys |  | 150 |  |  |  | 35 |  |
| thr | 17 | 106 | 56 |  |  | 48 |  |

Since the inverse matrix had to be square and number of amino acids were fifteen and the number of properties considered were nine the solutions of the matrix equation had to be divided into sets of nine and six as shown.

**Appendix 2**

Using matrices on nine of the fifteen amino acids and their properties, we find that the ratio of the nine coefficients $a_{ij}$ of the equation are as shown in case I below. The other six amino acid result are shown as case II below.

|  | I | II |
|---|---|---|
| dipole moment | 1 | 1 |
| ln partition coefficient | -0.062 | -0.056 |
| hydrophobicity | 0.13 | 0.13 |
| ln activity coefficient | -0.34 | -0.34 |
| solubility parameter | -0.002 |  |
| volume | 0.001 | 0.19 |
| CHI | -0.002 | -0.014 |
| molar refractivity | -0.011 | -0.053 |

Notice that the dipole moments, ln partition coefficients, hydrophobicity and ln activity coefficient are closely correlated, whereas the volume, CHI factor and molar refractivity are not.

**Figure Captions**

*Figure 1*   Linear Graphs of Sorted Merit of Nine Physical Properties for Fifteen Amino Acids

*Figure 2*   Radar Graph of Sorted Merit of. Nine Physical Properties for Fifteen Amino Acids

*Figure 3*   Linear Sorted Merit Graph of Three Solution Properties for Fifteen Amino Acids.

*Figure 4*   Linear Sorted Merit Graph of Four Electromagnetic Propertes for Fifteen Amino Acids

*Figure 5*   Linear Sorted Graph for the Hydrophobicity and Volume/20 for Fifteen Amino Acids

*Figure 6*   Sort of Sum of All Properties for Fifteen Amino Acids

*Figure 7*   Matrix Transpose Linear Graph of Nine properties for Nine Amino Acids

*Figure 8*   Matrix Transpose Linear Graph of Six Physical Properties for Six Amino Acids

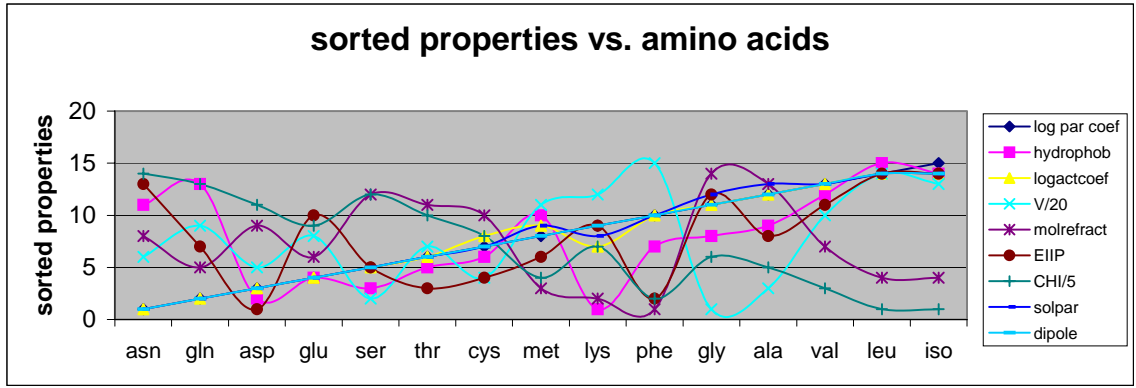

**Figure 1**

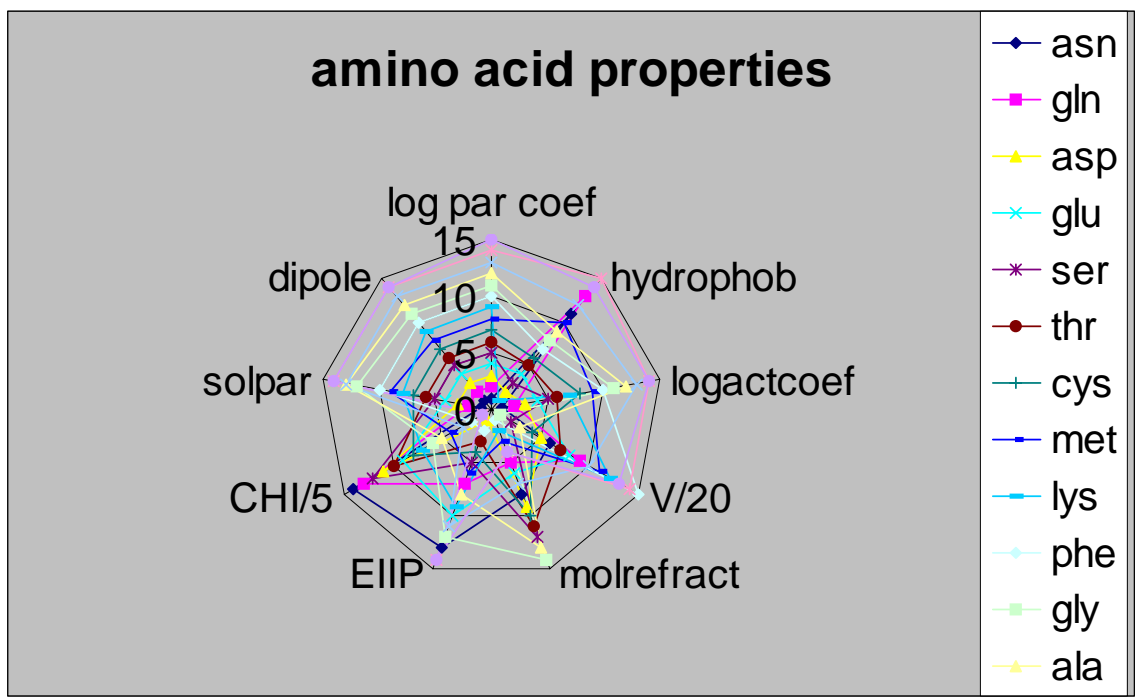

**Figure 2**

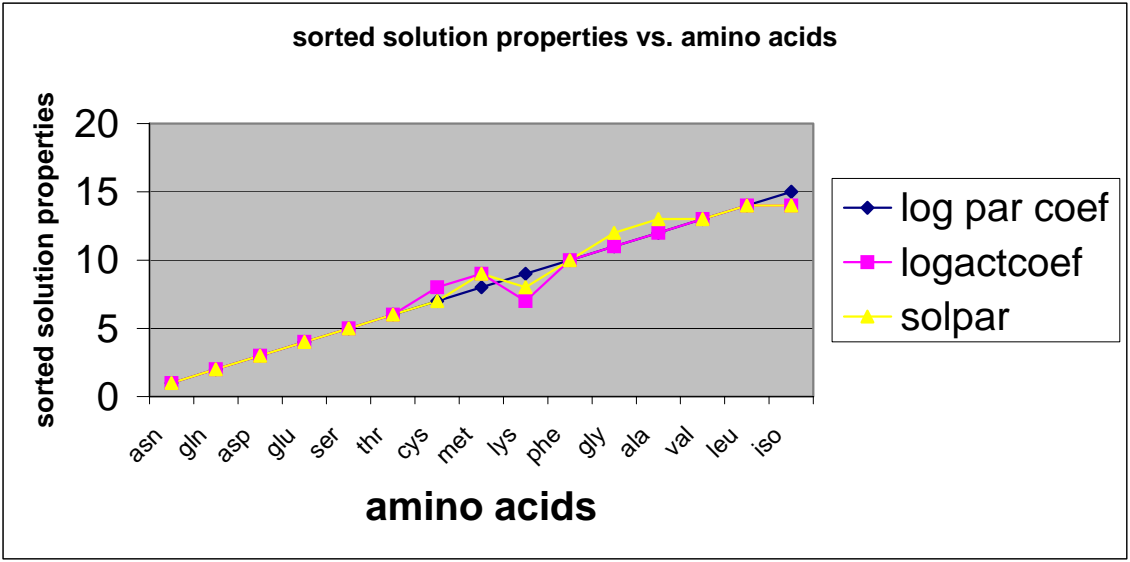

**Figure 3**

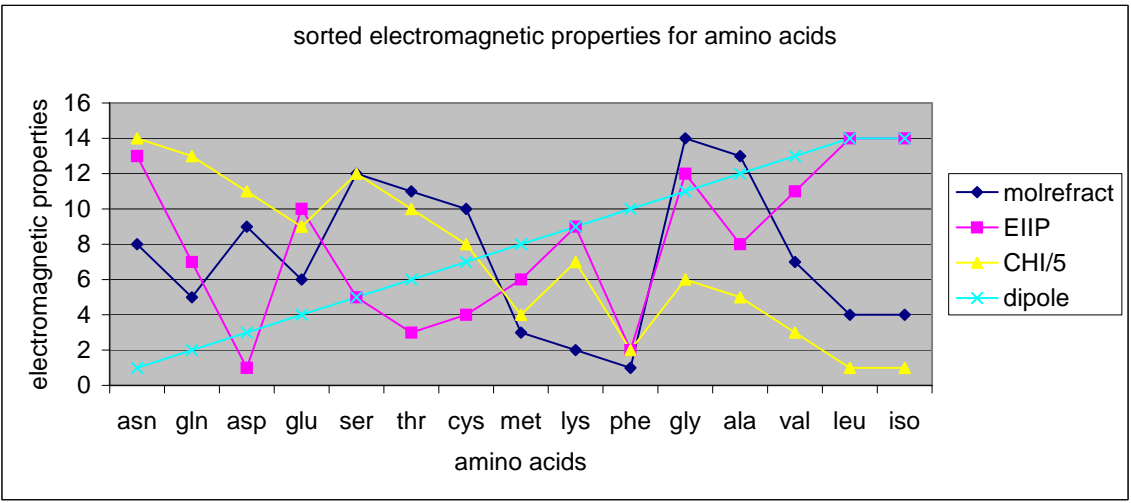

**Figure 4**

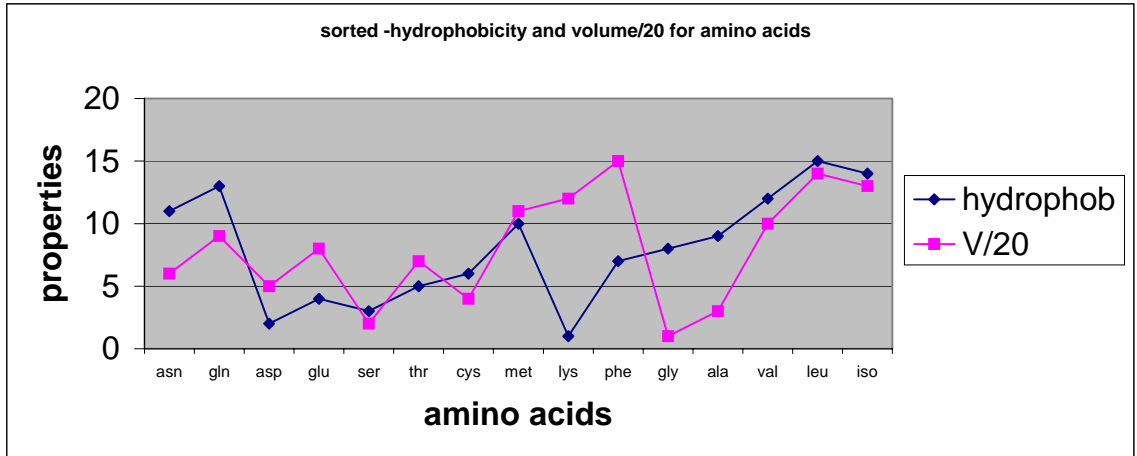

**Figure 5**

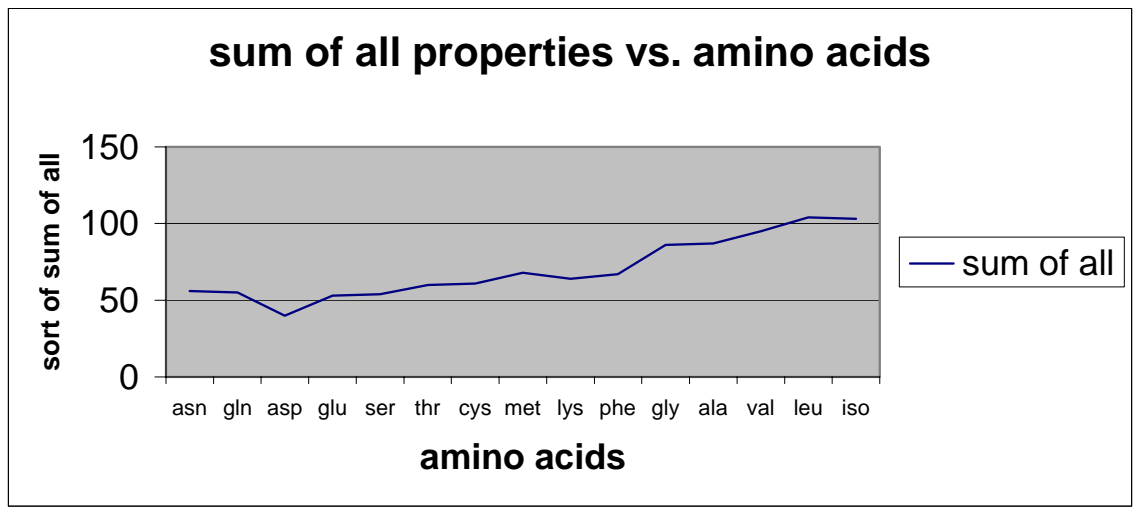

**Figure 6**

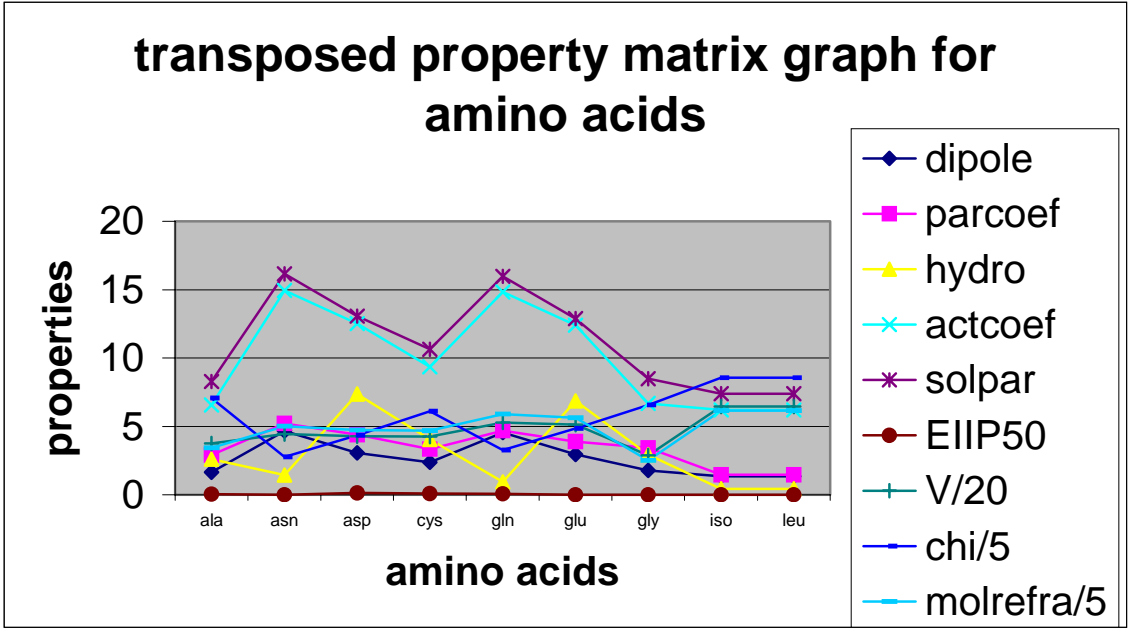

**Figure 7**

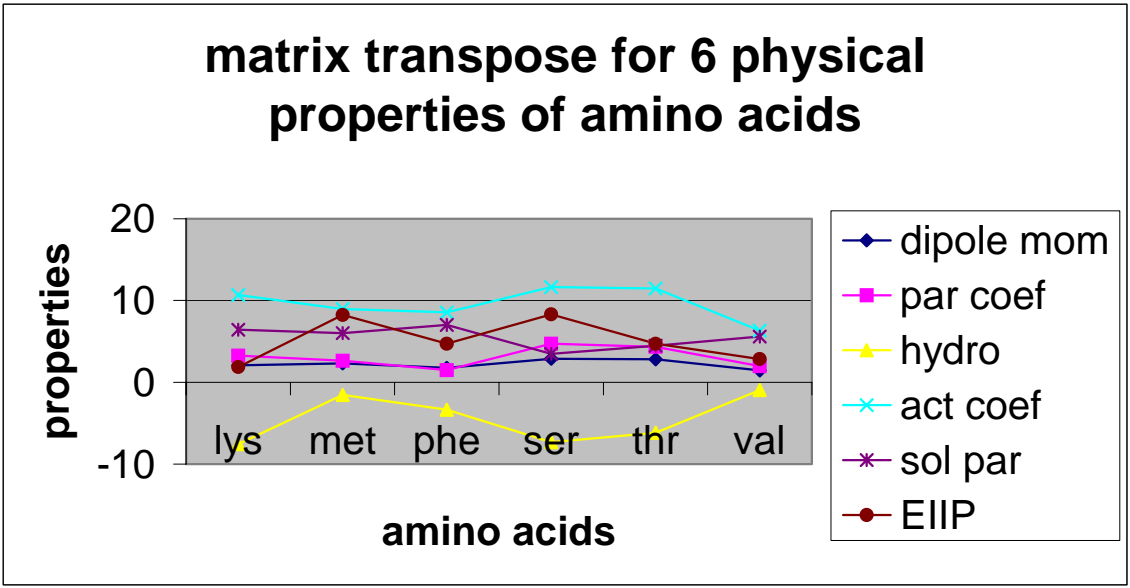

**Figure 8**